\documentclass[10pt,  letterpaper]{IEEEtran}
\usepackage{times,amsfonts,amssymb,amsmath,setspace}
\usepackage{color}
\usepackage{graphicx}
\usepackage{cite}
\usepackage{url}
\usepackage{bbm}\usepackage{mathrsfs}\usepackage{float}
\usepackage{lipsum}
\usepackage{multicol}

\usepackage{graphicx,url,bm,subfigure,comment}

\DeclareMathAlphabet\mathbfcal{OMS}{cmsy}{b}{n}
\newcommand{\insertfig}[4]{
\begin{figure}[ht]
\centerline{\includegraphics[width=#1\columnwidth]{#2.eps}}
\caption{#3}\label{#4}\end{figure}}

\DeclareMathAlphabet{\mathsfbf}{OT1}{cmss}{sbc}{n}

\newtheorem{proposition}{Proposition}[section]

\newcommand{\EE}{\mathbb{E}} 
\newcommand{\PP}{\mathbb{P}} 

\newcommand{\av}{{\bf a}}

\newcommand{\mv}{{\bf m}}
\newcommand{\nv}{{\bf n}}

\newcommand{\tv}{{\bf t}}

\newcommand{\vv}{{\bf v}}

\newcommand{\Am}{{\bf A}}

\newcommand{\Gm}{{\bf G}}

\newcommand{\Fb}{\mathbfcal F}
\newcommand{\Bb}{\mathbfcal B}

\newcommand{\Ac}{{\cal A}}
\newcommand{\Bc}{{\cal B}}
\newcommand{\Cc}{{\cal C}}
\newcommand{\Dc}{{\cal D}}

\newcommand{\Gc}{{\cal G}}
\newcommand{\Hc}{{\cal H}}
\newcommand{\Ic}{{\cal I}}

\newcommand{\Lc}{{\cal L}}

\newcommand{\Oc}{{\cal O}}

\newcommand{\alphav}{\boldsymbol{\alpha}}

\newcommand{\tauv}{\boldsymbol{\tau}}

\newcommand{\sgn}{\text{sgn}}

\def\ben{\begin{enumerate}}
\def\beq{\begin{equation}}
\def\beqa{\begin{eqnarray}}
\def\bit{\begin{itemize}}
\def\een{\end{enumerate}}
\def\eeq{\end{equation}}
\def\eeqa{\end{eqnarray}}
\def\eit{\end{itemize}}

\def\non{\nonumber\\}
\def\argmax{\mathop{\mathrm{arg~max}}\limits}

\begin{document}

\title{The Importance of Being Earnest \\ 
in Crowdsourcing Systems}

\author{A. Tarable, A. Nordio, E. Leonardi, and M. Ajmone Marsan
\IEEEcompsocitemizethanks{\IEEEcompsocthanksitem A. Tarable and A. Nordio are with CNR-IEIIT, Torino, Italy.\protect\\
E-mail: \{alberto.tarable; alessandro.nordio\}{@}ieiit.cnr.it
\IEEEcompsocthanksitem E. Leonardi and M. Ajmone Marsan are with the Department of Electronic and Telecommunications, Politecnico di Torino, Torino, Italy.\protect\\
E-mail: emilio.leonardi{@}polito.it, marco.ajmone{@}polito.it} %
\thanks{}}
\maketitle

\begin{abstract} 

This  paper presents  the first  systematic  investigation of  the potential  performance gains  for
crowdsourcing systems, deriving from available information  at the requester about individual worker
earnestness (reputation). In particular, we first formalize the optimal task assignment problem when
workers' reputation estimates are available, as the maximization of a monotone (submodular) function
subject to  Matroid constraints. Then, being  the optimal problem  NP-hard, we propose a  simple but
efficient  greedy  heuristic  task  allocation  algorithm.   We  also  propose  a  simple  ``maximum
a-posteriori`` decision rule. Finally, we test  and compare different solutions, showing that system
performance can greatly  benefit from information about workers' reputation.   Our main findings are
that: i) even  largely inaccurate estimates of  workers' reputation can be  effectively exploited in
the  task assignment  to greatly  improve system  performance; ii)  the performance  of the  maximum
a-posteriori decision rule  quickly degrades as worker reputation estimates  become inaccurate; iii)
when  workers' reputation  estimates  are  significantly inaccurate,  the  best  performance can  be
obtained by combining  our proposed task assignment  algorithm with the LRA decision  rule introduced in
the literature.

\end{abstract}

\section{Introduction}

Crowdsourcing is a term often adopted to identify networked systems that can be used for the
solution of a wide range of complex problems by integrating a large number of human and/or computer
efforts \cite{asurveyofcrowdsourcingsystems}.  Alternative terms, each one carrying its own specific
nuance, to identify similar types of systems are: collective intelligence, human computation,
master-worker computing, volunteer computing, serious games, voting problems, peer production,
citizen science (and others).  The key characteristic of these systems is that a {\em requester}
structures his problem in a set of {\em tasks}, and then assigns tasks to {\em workers} that provide
{\em answers}, which are then used to determine the correct task {\em solution} through a {\em
  decision} rule.  Well-known examples of such systems are SETI@home, which exploits unused computer
resources to search for extra-terrestrial intelligence, and the Amazon Mechanical Turk, which
allows the employment of large numbers of micro-paid workers for tasks requiring human intelligence
(HIT -- Human Intelligence Tasks). Examples of HIT are image classification, annotation, rating and
recommendation, speech labeling, proofreading, etc. In the Amazon Mechanical Turk, the workload
submitted by the requester is partitioned into several small atomic tasks, with a simple and
strictly specified structure. Tasks, which require small amount of work, are then assigned to
(human) workers. Since on the one hand answers may be subjective, and on the other task execution is
typically tedious, and the economic reward for workers is pretty small, workers are not 100 \%
reliable (earnest), in the sense that they may provide incorrect answers. Hence, the same task is
normally assigned in parallel (replicated) to several workers, and then a majority decision rule is
applied to their answers. A natural trade-off between the reliability of the decision and cost
arises; indeed, increasing the replication factor of every task, we can increase the reliability
degree of the final decision about the task solution, but we necessarily incur higher costs (or, for
a given fixed cost, we obtain a lower task throughput).  Although the pool of workers in
crowdsourcing systems is normally large, it can be abstracted as a finite set of shared resources,
so that the allocation of tasks to workers (or, equivalently, of workers to tasks) is of key
relevance to the system performance.

Some believe that crowdsourcing systems will provide a significant new type of work organization
paradigm, and will employ large numbers of workers in the future, provided that the main challenges
in this new type of organizations are correctly solved. In \cite{thefutureofcrowdwork} the authors
identify a dozen such challenges, including i) workflow definition and hierarchy, ii) task
assignment, iii) real-time response, iv) quality control and reputation. Task assignment and
reputation are central to this paper, where we discuss optimal task assignment with approximate
information about the quality of answers generated by workers (with the term worker reputation we
generally mean the worker earnestness, i.e., the credibility of a worker's answer for a given task,
which we will quantify with an error probability). Our optimization aims at minimizing the
probability of an incorrect task solution for a maximum number of tasks assigned to workers, thus
providing an upper bound to delay and a lower bound on throughput. A dual version of our
optimization is possible, by maximizing throughput (or minimizing delay) under an error probability
constraint. Like in most analyses of crowdsourcing systems, we assume no interdependence among
tasks, but the definition of workflows and hierarchies is an obvious next step. Both these issues
(the dual problem and the interdependence among tasks) are left for further work.
 
The performance of crowdsourcing systems is not yet explored in detail, and the only cases which
have been extensively studied in the literature assume that the quality of the answers provided by
each worker (the worker reputation) are not known at the time of task assignment. This assumption is
motivated by the fact that the implementation of reputation-tracing mechanisms for workers is
challenging, because the workers' pool is typically large and highly dynamical. Furthermore, in some
cases the anonymity of workers must be preserved. Nevertheless, we believe that a clear
understanding of the potential impact on the system performance of even approximate information
about the workers' reputation in the task assignment phase is extremely important, and can properly
assess the relevance of algorithms that trace the reputation of workers. Examples of algorithms that
incorporate auditing processes in a sequence of task assignments for the worker reputation
assessment can be found in \cite{antonio1,antonio2,antonio3,econ1,econ2,KDD09,ZSD10}.

Several algorithms were recently proposed in the technical literature to improve the performance of
crowdsourcing systems without a-priori information about worker reputation
\cite{Devavrat,shah2,shah3, adaptive, moderators}.  In particular, \cite{adaptive} proposed an
adaptive simple on-line algorithm to assign an appropriate number of workers to every task, so as to
meet a prefixed constraint on problem solution reliability. In~\cite{Devavrat,shah2,shah3, moderators},
instead, it was shown that the reliability degree of the final problem solution can be significantly
improved by replacing the simple majority decision rule with smarter decision rules that differently
weigh answers provided by different workers. Essentially the same decision strategy was
independently proposed in~\cite{Devavrat,shah2} and~\cite{moderators} for the case in which every
task admits a binary answer, and then recently extended in~\cite{shah3} to the more general
case. The proposed approach exploits existing redundancy and correlation in the pattern of answers
returned from workers to infer an a-posteriori reliability estimate for every worker. The derived
estimates are then used to properly weigh workers' answers.
 
The goal of this paper is to provide the first systematic analysis of the potential benefits
deriving from some form of a-priori knowledge about the reputation of workers. With this goal in
mind, first we define and analyze the task assignment problem when workers' reputation estimates are
available. We show that in some cases, the task assignment problem can be formalized as the
maximization of a monotone submodular function subject to Matroid constraints. A greedy
algorithm with performance guarantees is then devised.  In addition, we propose a simple ``maximum
a-posteriori`` (MAP) decision rule, which is well known to be optimal when perfect estimates of
workers' reputation are available.  Finally, our proposed approach is tested in several scenarios,
and compared to previous proposals.

Our main findings are:
\begin{itemize}
\item even largely inaccurate estimates of workers' reputation can be effectively exploited in the
  task assignment to greatly improve system performance;
\item the performance of the maximum a-posteriori decision rule quickly degrades as worker
  reputation estimates become inaccurate;
\item when workers' reputation estimates are significantly inaccurate, the best performance can be
  obtained by combining our proposed task assignment algorithm with the decision rule introduced in
  \cite{Devavrat, moderators}.
\end{itemize}

The rest of this paper is organized as follows.  Section \ref{sec:SA} presents and formalizes the
system assumptions used in this paper.  Section \ref{sec:PF} contains the formulation of the problem
of the optimal allocation of tasks to workers, with different possible performance objectives.
Section \ref{sec:allocation} proposes a greedy allocation algorithm, to be coupled with the MAP
decision rule described in Section \ref{sec:decision}.  Section \ref{sec:results} presents and
discusses the performance of our proposed approach in several scenarios, and compares it to those of
previous proposals.  Finally, Section \ref{sec:conclusions} concludes the paper and discusses
possible extensions.

\section{System Assumptions}
\label{sec:SA}

We consider $T$ binary tasks $\theta_1, \theta_2, \dots, \theta_T$, whose outcomes can be
represented by i.i.d. uniform random variables (RV's) $\tau_1, \tau_2, \dots, \tau_T$ over $\{\pm
1\}$, i.e., $ \PP \{\tau_t = \pm 1\} = \frac1{2}$, $t = 1,\dots, T$. In order to obtain a reliable
estimate of task outcomes, a requester assigns tasks to workers selected from a given population of
size $W$, by querying each worker $\omega_w$, $w = 1,\dots, W$ a subset of tasks.

Each worker is modeled as a binary symmetric channel (BSC) \cite[p. 8]{Cover}. This means that
worker $\omega_w$, if queried about task $\theta_t$, provides a wrong answer with probability $p_{t
  w}$ and a correct answer with probability $1-p_{t w}$. Note that we assume that the error
probabilities $p_{t w}$ depend on both the worker and the task, but they are taken to be
time-invariant, and generally unknown to the requester. The fact that the error probability may
depend, in general, both on the worker and the task reflects the realistic consideration that tasks
may have different levels of difficulty, that workers may have different levels of accuracy, and may
be more skilled in some tasks than in others.

Unlike the model in \cite{Devavrat,shah2}, we assume in this paper that, thanks to a-priori
information, the requester can group workers into classes, each one composed of workers with similar
accuracy and skills. In practical crowdsourcing systems, where workers are identified through
authentication, such a-priori information can be obtained by observing the results of previous task
assignments. More precisely, we suppose that each worker belongs to one of $K$ classes, $\Cc_1,
\Cc_2, \dots, \Cc_K$, and that each class is characterized, for each task, by a different
\emph{average} error probability, known to the requester. Let $\pi_{t k}$ be the average error
probability for class $\Cc_k$ and task $\theta_t$, $k = 1,\dots, K$, $t = 1,\dots, T$.  We emphasize
that $\pi_{t k}$ does not necessarily precisely characterize the reliability degree of individual
workers within class $k$ while accomplishing task $\theta_t$; this for the effect of possible
errors/inaccuracies in the reconstruction of user profiles. Workers with significantly different
degree of reliability can, indeed, coexist within class $k$. In particular our class
characterization encompasses two extreme scenarios:

\begin{itemize}
\item full knowledge about the reliability of workers, i.e., each worker belonging to class $\Cc_k$
  has error probability for task $\theta_t$ deterministically equal to $\pi_{t k}$, and
\item a hammer-spammer (HS) model~\cite{Devavrat}, in which perfectly reliable and completely
  unreliable users coexists within the same class.  A fraction $2\pi_{t k}$ of workers in class
  $\Cc_k$, when queried about task $\theta_t$, has error probability equal to $\frac1{2}$ (the
  spammers), while the remaining workers have error probability equal to zero (the hammers).
\end{itemize}

Suppose that class $\Cc_k$ contains a total of $W_k$ workers, with $W = \sum_{k=1}^K W_k$.  The
first duty the requester has to carry out is the assignment of tasks to workers. We impose the
following two constraints on possible assignments:
\begin{itemize}
\item a given task $\theta_t$ can be assigned at most once to a given worker $\omega_w$, and
\item no more than $r_w$ tasks can be assigned to worker $\omega_w$.
\end{itemize} 
Notice that the second constraint arises from practical considerations on the amount of load a
single worker can tolerate. We also suppose that each single assignment of a task to a worker has a
\emph{cost}, which is independent of the worker's class.  In practical systems, such cost represents
the (small) wages per task the requester pays the worker, in order to obtain answers to his
queries. Alternatively, in voluntary computing systems, the cost can describe the time necessary to
perform the computation.  The reader may be surprised by the fact that we assume worker cost to be
independent from the worker class, while it would appear more natural to differentiate wages among
workers, favoring the most reliable, so as to incentivize workers to properly
behave~\cite{econ1,econ2}.  Our choice, however, is mainly driven by the following two
considerations: i) while it would be natural to differentiate wages according to the individual
reputation of workers, when the latter information is sufficiently accurate, it is much more
questionable to differentiate them according to only an average collective reputation index, such as
$\pi_{t k}$, especially when workers with significantly different reputation coexist within the same
class; ii) since in this paper our main goal is to analyze the impact on system performance of
a-priori available information about the reputation of workers, we need to compare the performance
of such systems against those of systems where the requester is completely unaware of the worker
reputation, under the same cost model. Finally, we wish to remark that both our problem formulation
and proposed algorithms naturally extend to the case in which costs are class-dependent.

Let an \emph{allocation} be a set of assignments of tasks to workers.  More formally, we can
represents a generic allocation with a set $\Gc$ of pairs $(t,w)$ with $t \in \{1,\cdots, T\}$ and
$w \in \{1,\cdots,W\}$, where every element $(t,w)\in\Gc$ corresponds to an individual task-worker
assignment.  Let $\Oc$ be the complete allocation set, comprising every possible individual
task-worker assignment (in other words $\Oc$ is the set composed of all the possible $T\cdot W$
pairs $(t,w)$). Of course, by construction, for any possible allocation $\Gc$, we have that $\Gc
\subseteq \Oc$. Hence, the set of all possible allocations corresponds to the power set of $\Oc$,
denoted as $2^{\Oc}$.

The set $\Gc$ can also be seen as the edge set of a bipartite graph where the two node subsets
represent tasks and workers, and there is an edge connecting task node $t$ and worker node $w$ if
and only if $(t,w) \in \Gc$. It will be sometimes useful in the following to identify the allocation
with the biadjacency matrix of such graph. Such binary matrix of size $T \times W$ will be denoted
$\Gm = \{g_{tw}\}$ and referred to as the \emph{allocation matrix}. In the following we will
interchangeably use the different representations, according to convenience.

In this work, we suppose that the allocation is non-adaptive, in the sense that all assignments are
made before any decision is attempted. With this hypothesis, the requester must decide the
allocation only on the basis of the a-priori knowledge on worker classes. Adaptive allocation
strategies can be devised as well, in which, after a partial allocation, a decision stage is
performed, and gives, as a subproduct, refined a-posteriori information both on tasks and on
workers' accuracy. This information can then be used to optimize further
assignments. However, in \cite{shah2} it was shown that non-adaptive allocations are order
optimal in a single-class scenario.

When all the workers' answers are collected, the requester starts deciding, using the received
information. Let $\Am = \{a_{tw}\}$ be a $T \times W$ random matrix containing the workers' answers
and having the same sparsity pattern as $\Gm$. Precisely, $a_{tw}$ is nonzero if and only if
$g_{tw}$ is nonzero, in which case $a_{tw} = \tau_t$ with probability $1-p_{tw}$ and $a_{tw} =
-\tau_t$ with probability $p_{tw}$. For every instance of the matrix $\Am$ the output of the
decision phase is an estimate $\hat{\tau}_1, \hat{\tau}_2, \dots, \hat{\tau}_T$ for task values.

\section{Problem Formulation\label{sec:PF}}
In this section, we formulate the problem of the optimal allocation of tasks to workers, with
different possible performance objectives. We formalize such problem under the assumption that each
worker in class $\Cc_k$ has error probability for task $\theta_t$ deterministically equal to
$\pi_{tk}$.

If the individual error probability of the workers within one class is not known to the scheduler, it
becomes irrelevant which worker in a given class is assigned the task. What only matters is actually
how many workers of each class is assigned each task. By sorting the columns (workers) of the
allocation matrix $\Gm$, we can partition it as \beq \Gm = \left[ \Gm_1, \Gm_2, \dots, \Gm_K \right]
\eeq where $\Gm_k$ is a binary matrix of size $T \times W_k$ representing the allocation of tasks to
class-$k$ workers. Define $W^{(k)} = \sum_{i=1}^k W_{i}$ and $W^{(0)}=0$. Define also $d_{tk}$ as
the weight (number of ones) in the $t$-th row of matrix $\Gm_k$, which also represents the degree of
the $t$-th task node in the subgraph containing only worker nodes from the $k$-th class.

\subsection{Optimal allocation}
We formulate the problem of optimal allocation of tasks to workers as a combinatorial optimization
problem for a maximum overall cost. Namely, we fix the maximum number of assignments (or,
equivalently, the maximum number of ones in matrix $\Gm$) to a value $C$, and we seek the best
allocation in terms of degree set $\Dc = \{ d_{tk}, t = 1, 2, \dots,T, k = 1, 2, \dots,
K\}$. Let $P(\Dc)$ be a given performance parameter to be maximized. Then, the problem can be
formalized as follows.
\begin{eqnarray}  
\Dc^{\mathrm{opt}} 
 &=& \argmax_{\Dc}\,\, P(\Dc) \non
 &\mbox{s.t.}&  d_{tk} \,\,\mathrm{integer},\,\,\, 0 \leq d_{tk} \leq W_k, \,\,\,t=1,2,\ldots,T,\non  
 &&   \;\;\;\;\;\;\;\;\;\;\;\;\;\;\;\;\;\;\;\;\;\;\;\;\;\;\;\;\;\;\;\;\;\;\;\;\;\;\;\;\;\;\;\;\;k=1,2,\dots,K \non 
 && \sum_{t=1}^T d_{tk} \leq \sum_{w=W^{(k-1)}+1}^{W^{(k)}} r_w, \,\,\,k=1,2,\dots,K, \non
 && \sum_{t=1}^T \sum_{k=1}^K d_{tk} \leq C 
\label{eq:optimal_allocation}
\end{eqnarray}
where the first constraint expresses the fact that $d_{tk}$ is the number of ones in the $t$-th row
of $\Gm_k$, the second constraint derives from the maximum number of tasks a given worker can be
assigned, and the third constraint fixes the maximum overall cost.

Note that it could also be possible to define a dual optimization problem, in which the optimization
aims at the minimum cost, subject to a maximum admissible error probability; this alternative problem
is left for future work.

By adopting the set notation for allocations, we can denote with $\Fb$ the family of all feasible
allocations (i.e.  the collection of all the allocations respecting the constraints on the total
cost and the worker loads).  Observe that by construction $\Fb \subseteq 2^{\Oc}$ is composed of all
the allocations $\Gc$ satisfying: i) $|\Gc|\le C$, and ii) $|\Lc(w,\Gc)|\le r_w$ $\forall w$, where
$\Lc(w,\Gc)$ represents the set of individual assignments in $\Gc$ associated to $w$.  The advantage
of the set notation is that we can characterize the structure of the family $\Fb$ on which the
performance optimization must be carried out; in particular, we can prove that:
\begin{proposition} \label{prop-Matroid}
The family $\Fb$ forms a Matroid~\cite{Calinescu}. Furthermore, $\Fb$ satisfies the following
property.  Let $\Bb \in \Fb$ be the family of maximal sets in $\Fb$, then
\[ q= \frac{\max_{\Gc \in \Bb}  |\Gc|}{\min_{\Gc \in \Bb}  |\Gc|}=1\,.\]
\end{proposition}
The proof is reported in Appendix  \ref{app:matroid} along with the definition of a Matroid.

\subsubsection{Computational  complexity}
the complexity of the above optimal allocation problem heavily depends on the structure of the
objective function $P(\Dc)$ (which is rewritten as $P(\Gc)$ when we adopt the set notation). As a
general property, observe that necessarily $P(\Gc)$ is monotonic, in the sense that $P(\Gc_1) \le
P(\Gc_2)$ whenever $\Gc_1 \subset \Gc_2$. However, in general, we cannot assume that $P(\Gc)$
satisfies any other specific property (some possible definitions for $P(\Gc)$ are given next).  For
a general monotonic objective function, the optimal allocation of tasks to workers can be shown to
be NP-hard, since it includes as a special case the well-known problem of the maximization of a
monotonic submodular function, subject to a uniform Matroid constraint
(see~\cite{Calinescu})\footnote{A set function $f: 2^{\Oc}\to \mathbb{R}^+ $ is said to be
  submodular if: $\forall \Ac,\Bc \in : 2^{\Oc}$ we have $f(\Ac \cup \Bc) + f(\Ac \cap \Bc) \le
  f(\Ac) + f(\Bc)$. The problem of the maximization of a monotonic submodular function subject to a
  uniform Matroid constraint corresponds to: \{$\max_{|\Ac|\le K} f(\Ac)$ for $K<|\Oc|$ with $f(.)$
  submodular.\}}.  When $P(\Gc)$ is submodular, the optimal allocation problem falls in the
well-known class of problems related to the maximization of a monotonic submodular function subject
to Matroid constraints. For such problems, it has been proved that a greedy algorithm yields a
1/(1+$q$)-approximation~\cite{Calinescu} (where $q$ is defined as in Proposition
\ref{prop-Matroid}).

In the next subsections, we consider different choices for the performance parameter $P(\Dc)$.
\subsection{Average task error probability }

A possible objective of the optimization, which is most closely related to typical performance
measures in practical crowdsourcing systems, is the average task error probability, which is defined
as: 
\beq 
P_1(\Dc) = -\frac1{T} \sum_{t=1}^{T} P_{e,t} \eeq with $P_{e,t} = \PP \{\hat{\tau}_t
\neq \tau_t \} = \PP \{\hat{\tau}_t \neq 1 | \tau_t=1\}$ 
where the second equality follows from
symmetry.  Of course, $P_{e,t}$ can be exactly computed only when the true workers' error
probabilities $p_{tw}$ are available; furthermore it heavily depends on the adopted decoding
scheme. As a consequence, in general, $P_{e,t}$ can only be approximately estimated by the requester
by confusing the actual worker error probability $p_{tw}$ (which is unknown) with the corresponding
average class error probability $\pi_{tk}$.  Assuming a maximum-a-posteriori (MAP) decoding scheme,
namely, $\hat{\tau}_t(\alphav) = \arg \max_{\tau_t \in \pm 1} \PP \{\tau_t |\av_{t}= \alphav \}$,
where $\av_t$ is the $t$-th row of $\Am$ and $\alphav$ is its observed value, we have 
\beq \label{eq:error_probability_task_i}
P_{e,t} = \sum_{\alphav: \PP \{\tau_t =1|\av_{t}= \alphav \} < 1/2} \PP \{\av_{t} = \alphav|\tau_t=1
\} .  
\eeq 
It is easy to verify that the exact computation of the previous average task error
probability estimate requires a number of operations growing exponentially with the number of
classes $K$.  Thus, when the number of classes $K$ is large, the evaluation of
(\ref{eq:error_probability_task_i}) can become critical.

To overcome this problem, we can compare the performance of different allocations on the basis of a
simple pessimistic estimate of the error probability, obtained by applying the Chernoff bound to the
random variable that is driving the maximum-a-posteriori (MAP) decoding (details on a MAP decoding
scheme are provided in the next section). We have:
\[
P_{e,t} \le \hat{P}_{e,t}=\exp\left(- \frac{ \sum_k d_{tk}(1- 2 \pi_{tk})z_{tk}}{ \sum_j (d_{tk} z_{tk})^2}\right)
\]
where $z_{tk}=\log(\frac{1-\pi_{tk}}{ \pi_{tk}})$. Thus, the performance metric associated with an allocation becomes:
\[
 P_2(\Dc)= -\frac1{T} \sum_{t=1}^{T} \hat{P}_{e,t} 
\]
The computation of $P_2(\Dc)$ requires a number of operations that scales linearly with the product
$T\cdot K$.  At last, we would like to remark that in practical cases we expect the number of
classes to be sufficiently small (order of few units), in such cases the evaluation of
(\ref{eq:error_probability_task_i}) is not really an issue.

\subsection{Overall mutual information\label{sec:mutual_info}}
An alternative information-theoretic choice for $P(\Dc)$ is the mutual information between the
vector of RVs associated with tasks $\tauv = (\tau_1, \tau_2,\dots, \tau_T)$ and the answer matrix
$\Am$, i.e., 
\beq
\label{eq:mutual_info}
P_3(\Dc) = I(\Am; \tauv) = \sum_{t=1}^T I(\av_{t}; \tau_t)\,.  
\eeq 
It is well known that a tight
relation exists between the mutual information and the achievable error probability, so that a
maximization of the former corresponds to a minimization of the latter.  We remark, however, that,
contrary to error probability, mutual information is independent from the adopted decoding
scheme, because it refers to an optimal decoding scheme. This property makes the adoption of the
mutual information as the objective function for the task assignment quite attractive, since it
permits to abstract from the decoding scheme.  The second equality in (\ref{eq:mutual_info}) comes
from the fact that tasks are independent and workers are modeled as BSCs with known error
probabilities, so that answers to a given task do not give any information about other tasks. By
definition \beq
\label{eq:mutual_info_definition}
I(\av_{t}; \tau_t) = H(\av_{t}) - H(\av_{t} | \tau_t) = H(\tau_t) - H(\tau_t |\av_{t} ) 
\eeq 
where $H(a)$ denotes the entropy of the RV $a$, given by
\[ H(a) = -\EE_a [\log \PP(a)] \]  
and for any two random variables $a,b$, $H(a|b)$ is the conditional entropy defined as
\[ H(a|b) = -\EE_b\EE_{a|b} [\log \PP(a|b)]. \]  
In what follows, we assume perfect knowledge of worker reliabilities, i.e., we assume that each
class-$k$ worker has error probability with respect to task $\tau_t$ exactly equal to $\pi_{tk}$,
remarking than in the more general case, the quantities we obtain by substituting $p_{tw}$ with the
corresponding class average $\pi_{tk}$, can be regarded as computable approximations for the true
uncomputable mutual information.

Since we have modeled all workers as BSCs, each single answer is independent of everything else given the task value, so that
\beq \label{eq:conditional_entropy_A_given_t}
H(\av_{t} | \tau_t) = \sum_{a_{tw} \neq 0} H(a_{tw} | \tau_t) = \sum_{k=1}^K d_{tk} H_b(\pi_{tk}) .
\eeq
where
\[
H_b(p) = -p\log p -(1-p) \log (1-p).
\]
For the second equality in \eqref{eq:mutual_info_definition}, $H(\tau_t)=1$ because $\tau_t$ is a
uniform binary RV, and
\begin{eqnarray}
H(\tau_t |\av_{t} ) &=& \sum_{\alphav} \PP \{\av_{t} = \alphav\} H(\tau_t |\av_{t}= \alphav  ) \nonumber \\
&=& \sum_{\alphav} \PP \{\av_{t} = \alphav\} H_b(\PP \{\tau_t =1|\av_{t}= \alphav \}  ) 
\end{eqnarray}
where $\alphav$ runs over all possible values of $\av_{t}$.

By symmetry, for every $\alphav$ such that $\PP \{\tau_t =1|\av_{t}= \alphav \} < \frac1{2} $, there
is $\alphav'$ such that $\PP \{\av_{t} = \alphav'\} = \PP \{\av_{t} = \alphav\} $ and $\PP \{\tau_t
=1|\av_{t}= \alphav' \} = 1 - \PP \{\tau_t =1|\av_{t}= \alphav \}$.  As a consequence, we can write
\begin{eqnarray}
H(\tau_t | \av_{t} ) 
&=& \!\!\!\! 2 \!\!\!\!\!\!\!\!  \sum_{\alphav: \PP \{\tau_t =1|\av_{t}= \alphav \} < 1/2} \hspace{-5ex}\PP \{\av_{t} = \alphav\} H_b(\PP \{\tau_t =1|\av_{t}= \alphav \}) \nonumber \\
&\hspace{-5ex} = &\hspace{-12ex} \sum_{\alphav:  \PP \{\tau_t =1|\av_{t}= \alphav \} < 1/2} 
\hspace{-7ex}\left( \PP \{\av_{t} = \alphav|\tau_t=1 \} +\PP \{\av_{t} = \alphav|\tau_t=-1 \} \right) \cdot \nonumber \\
& & \hspace{-4ex}H_b(\PP \{\tau_t =1|\av_{t}= \alphav \}) 
 \label{eq:conditional_entropy_task_given_answers}
\end{eqnarray}
Notice the relationship of the above expression with \eqref{eq:error_probability_task_i}. If in
\eqref{eq:conditional_entropy_task_given_answers} we substitute $H_b(\PP \{\tau_t =1|\av_{t}=
\alphav \})$ with $\PP \{\tau_t =1|\av_{t}= \alphav \}$, thanks to Bayes' rule, we obtain
\eqref{eq:error_probability_task_i}.

An explicit computation of $I(\Am; \tauv)$ can be found in Appendix \ref{app:mutual}. As for the task error
probability, the number of elementary operations required to compute $I(\Am; \tauv)$ grows
exponentially with the number of classes $K$.

An important property that mutual information satisfies is submodularity. This property provides
some guarantees about the performance of the greedy allocation algorithm described in
Section~\ref{sec:greedy}.

\begin{proposition}[Submodularity of the mutual information]
\label{prop-submodularity}
Let $\Gc$ be a generic allocation for task $\theta$. Then, the mutual information $I(\Gc; \theta)$ is a
submodular function.
\end{proposition}
\begin{IEEEproof}
The proof is given in Appendix~\ref{app:submodularity}
\end{IEEEproof}

\subsection{Max-min performance parameters}

The previous optimization objectives represent a sensible choice whenever the target is to optimize
the {\em average} task performance. However, in a number of cases it can be more appropriate to
optimize the worst performance among all tasks, thus adopting a max-min optimization approach.

Along the same lines used in the definition of the previous optimization objectives, we can obtain three other possible choices of performance parameters to be used in the optimization problem defined in \eqref{eq:optimal_allocation}, namely, the maximum task error probability,
\beq
P_4(\Dc) = - \max_{t=1,\dots,T} P_{e,t} 
\eeq
the Chernoff bound on the maximum task error probability,
\beq
P_5(\Dc) = - \max_{t=1,\dots,T} \hat{P}_{e,t} 
\eeq
and the minimum mutual information,
\beq
P_6(\Dc) = \min_{t=1, 2, \dots, T} I(\av_{t}; \tau_t).
\eeq

\section{Allocation strategies}
\label{sec:allocation}

As we observed in Section \ref{sec:PF}, the optimization problem stated in
\eqref{eq:optimal_allocation} is NP-hard, but the submodularity of the mutual information objective
function over a Matroid, coupled with a greedy algorithm yields a 1/2-approximation~\cite{Calinescu}
(see Proposition \ref{prop-Matroid}). We thus define in this section a greedy task assignment
algorithm, to be coupled with the MAP decision rule which is discussed in the next section.

\subsection{Greedy task assignment\label{sec:greedy}}
The task assignment we propose to approximate the optimal performance is a simple greedy algorithm
that starts from an empty assignment ($\Gc^{(0)}= \emptyset$), and at every iteration $i$ adds to
$\Gc^{(i-1)}$ the individual assignment $(t,w)^{(i)}$, so as to maximize the objective function. In
other words;
\[
(t,w)^{(i)}= \argmax_{(t,w) \in \Oc\setminus \Gc^{(i-1)},(\Gc^{(i-1)}\cup \{(t,w)\})\in \Fb } \hspace{-3ex}P(\Gc^{(i-1)}\cup\{(t,w)\})
\]
The algorithm stops when no assignment can be further added to $\Gc$ without violating some
constraint.

To execute this greedy algorithm, at step $i$, for every task
$t$, we need to i) find, if any, the best performing worker to which task $t$ can be assigned
without violating constraints, and mark the assignment $(t,w)$ as a candidate assignment; ii)
evaluate for every candidate assignment the performance index $P(\Gc^{(i-1)}\cup(t,w))$ $\forall
t$; iii) choose among all the candidate assignments the one that greedily optimizes performance.

Observe that, as a result, the computational complexity of our algorithm is $O(T^2\cdot W Z)$ where
$Z$ represents the number of operations needed to evaluate $P(\Gc)$.

Note that in light of both Propositions~\ref{prop-Matroid} and~\ref{prop-submodularity}, the above
greedy task assignment algorithm provides a $1/2$-approximation when the objective function
$P_3(\Gc)$ is chosen.  Furthermore, we wish to mention that a better $1-1/e$-approximation can be
obtained by cascading the above greedy algorithm with the special local search optimization
algorithm proposed in~\cite{Calinescu}; unfortunately, the corresponding cost in terms of
computational complexity is rather severe, because the number of operations requested to run the
local search procedure is $\widetilde{O}((T\cdot W)^8Z)$
\footnote{The function $f(n)$ is $\widetilde{O}(g(n))$ if $f(n) =O(g(n)\log^bn)$ for any positive
  constant $b$.}.

\subsection{Uniform allocation}
Here we briefly recall that \cite{Devavrat,shah2} proposed a simple task allocation strategy (under
the assumption that workers are indistinguishable) according to which a random regular bipartite
graph is established between tasks and selected workers. Every selected worker is assigned the
same maximal number of tasks, i.e. $r_w=r$ $\forall w$, except for rounding effects induced by the
constraint on the maximum total number of possible assignments $C$.

\section{Decision rules\label{sec:decision}}

\subsection{Majority voting}
Majority voting is the simplest possible task-decision rule which is currently implemented in all
real-world crowdsourcing systems.  For every task $\theta_t$, it simply consists in counting the
$\{+1\}$ and the $\{-1\}$ in $\av_t$ and then taking $\hat{\tau}_t(\av_t)$ in accordance to the
answer majority.  More formally: 
\beq \label{eq:Majority_decision_rule} 
\hat{\tau}_t(\av_t) = \mathrm{sgn} \left( \sum_w a_{tw} \right).  
\eeq

\subsection{MAP decision rule}
We investigate the performance of the greedy task assignment algorithm, when coupled with the
MAP decision rule for known workers reputation.

Given an observed value of $\av_t$, the posterior log-likelihood ratio (LLR) for task $\tau_t$ is
\begin{eqnarray}
\mathrm{LLR}_t(\av_t) & = & \log \frac{\PP \{\tau_t = 1 | \av_t\}}{\PP \{\tau_t = -1 | \av_t\}} \non
& = & \log \frac{\PP \{\av_t | \tau_t = 1\} }{\PP \{\av_t | \tau_t = -1\} } \non 
& = & \sum_{w: a_{tw} \neq 0} \log \frac{\PP \{a_{tw} | \tau_t = 1\} }{\PP \{a_{tw} | \tau_t = -1\} } 
\end{eqnarray}
where the second equality comes from Bayes' rule and the fact that tasks are uniformly distributed
over $\pm 1$, and the third equality comes from modelling workers as BSCs.
Let $m_{tk}$ be the number of ``$-1$'' answers to task $t$ from class-$k$ workers. Then
\beq \label{eq:LLRaposteriori}
\mathrm{LLR}_t(\av_t) = \sum_{k=1}^K \left(d_{tk} - 2 m_{tk}\right) \log \frac{1-\pi_{tk}}{\pi_{tk}}.
\eeq

The MAP decision rule outputs the task solution estimate $\hat{\tau}_t=1$ if $\mathrm{LLR}_t > 0$
and $\hat{\tau}_t=-1$ if $\mathrm{LLR}_t < 0$, that is, \beq \label{eq:MAP_decision_rule}
\hat{\tau}_t(\av_t) = \mathrm{sgn} \left( \mathrm{LLR}_t(\av_t) \right).  \eeq

Observe that the computation of \eqref{eq:LLRaposteriori} has a complexity growing only linearly
with $K$, and that, according to \eqref{eq:MAP_decision_rule}, each task solution is estimated
separately.  Note also that, whenever worker reputation is \emph{not} known a-priori, the above
decision rule is no more optimal, since it neglects the information that answers to other tasks can
provide about worker reputation.

\subsection{Low-rank approximation (LRA)}
Finally, for the sake of comparison, we briefly recall here the Low-Rank Approximation decision rule
proposed in~\cite{Devavrat,shah2,moderators} for the case when: i) no a-priori information about the
reputation of workers is available, ii) the error probability of every individual worker $w$ is the
same for every task, i.e., $p_{tw}=p_w$ $\forall t$.  The LRA decision rule was shown to provide
asymptotically optimal performance under assumptions i) and ii)~\cite{shah2}.

Denote with $\vv$ the leading right singular vector of $\Am$, the LRA decision is taken according
to:
\[   \hat{\tau}_t(\av_t)=\sgn\left(\mathrm{LRA}(\av_t)\right) \]
where
\[ \mathrm{LRA}(\av_t)=  \sum_w a_{tw} v_w \]
The idea underlying the LRA decision rule is that each component of the leading singular vector of
$\Am$, measuring the degree of coherence among the answers provided by the correspondent worker,
represents a good estimate of the worker reputation.

\begin{figure*}[ht]
 \centering
 \subfigure[]
   {\includegraphics[width=0.32\textwidth]{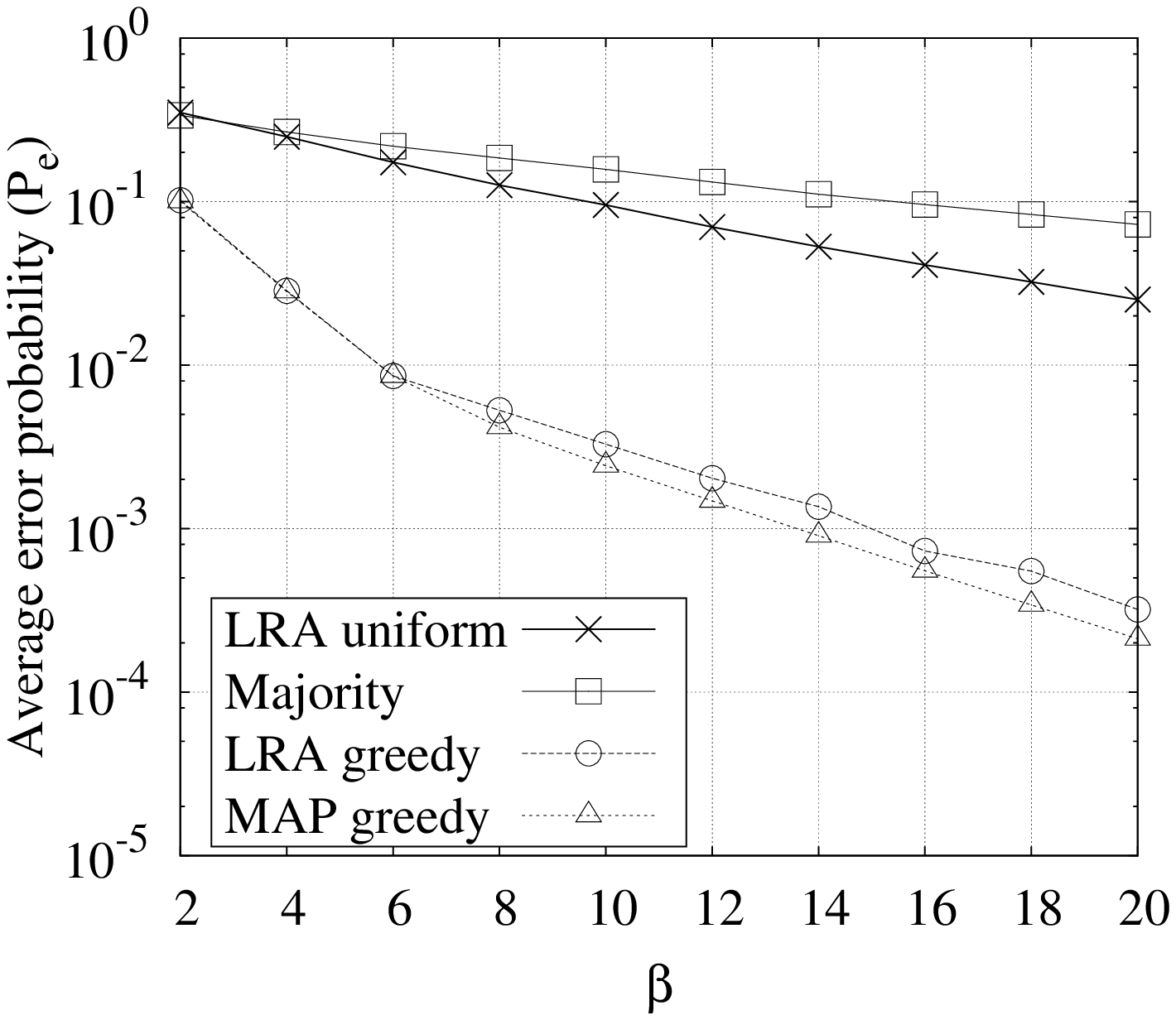}}
 \subfigure[]
   {\includegraphics[width=0.32\textwidth]{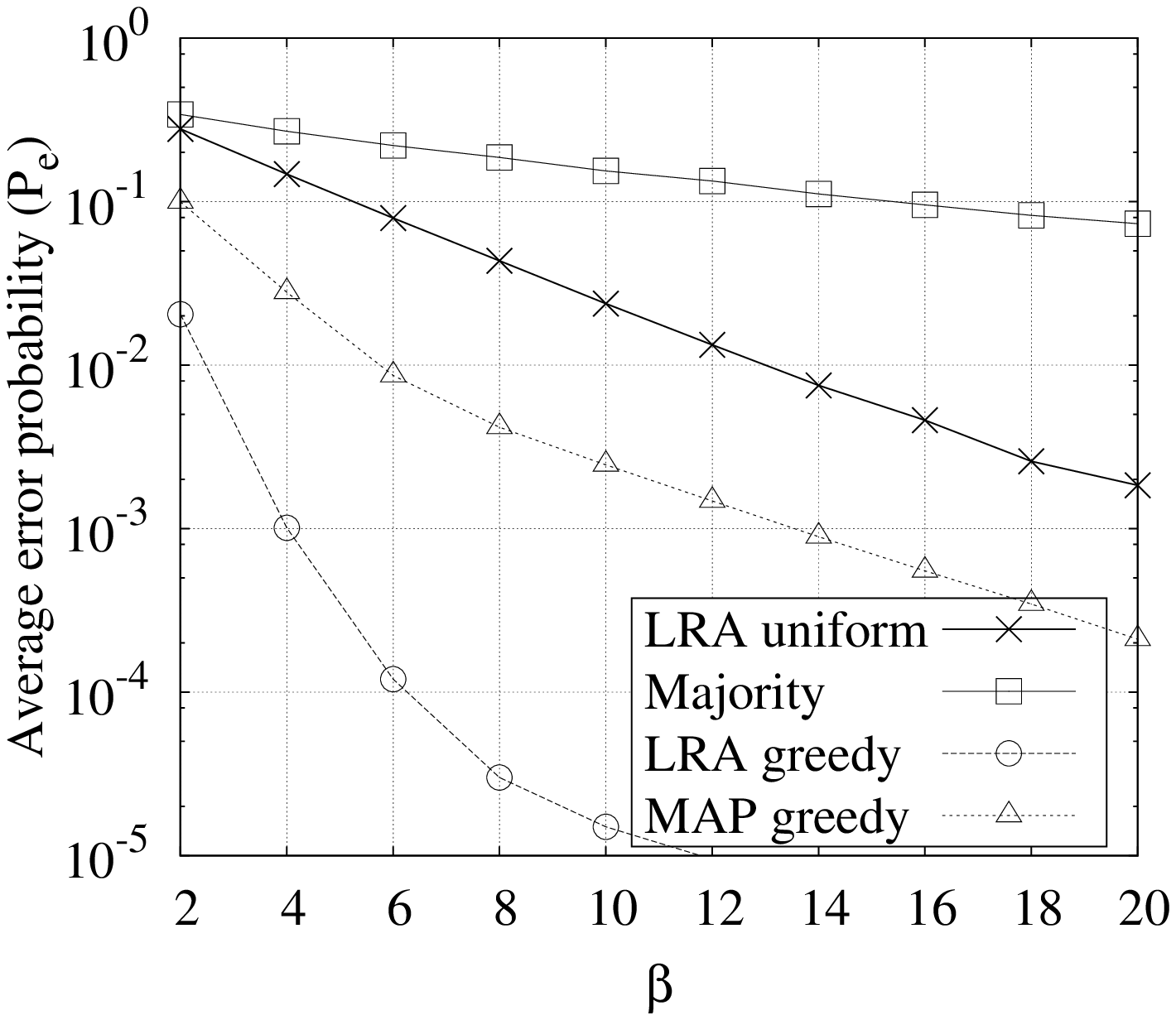}}
 \subfigure[]
   {\includegraphics[width=0.32\textwidth]{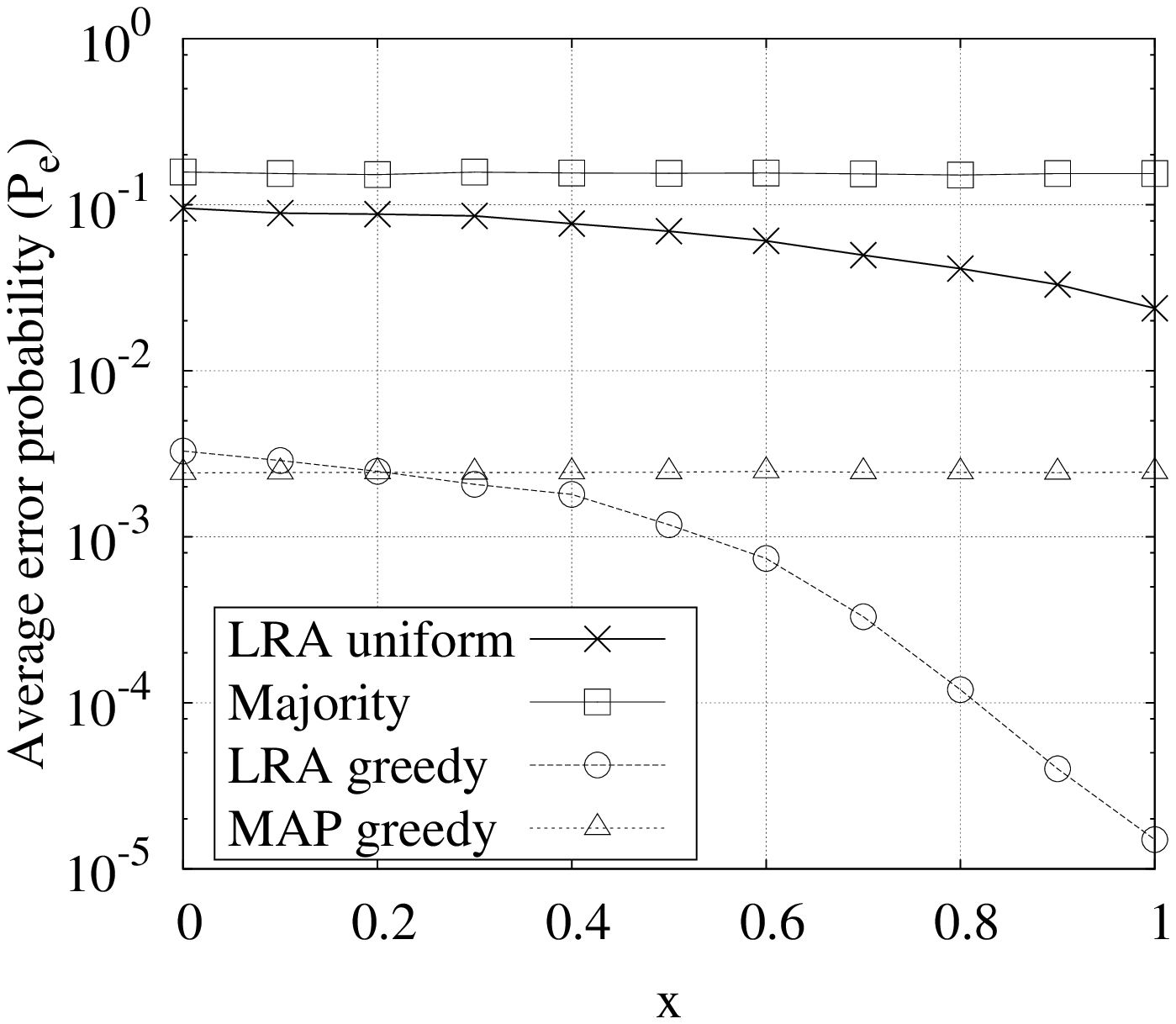}}
 \caption{Average error probability as a function of the average number of workers per task, $\beta$, and
   of the parameter $x$, for $\pi_{t1}=0.1,\pi_{t2}=0.2,\pi_{t3}=0.5$, $W_1 = 30, W_2 = 120$, and
   $W_3 = 150$. In figure (a) and (b) we set $x=0$ and $x=1$, respectively, while in figure (c) we set $\beta=10$.}
 \label{fig:figure1}
 \end{figure*}


\section{Results}
\label{sec:results}

In this section, we study the performance of a system where $T=100$ tasks are assigned to a set of
workers which are organized in $K=3$ classes.  Each worker can handle up to 20 tasks, i.e.,
$r_w=20$, $w=1,\ldots, W$.

We compare the performance of the allocation algorithms and decision rules described
in Sections~\ref{sec:allocation} and ~\ref{sec:decision}, in terms of achieved average error
probability, $P_e$.  More specifically, we study the performance of:
\begin{itemize}
\item the ``Majority voting'' decision rule applied to the ``Uniform allocation'' strategy, hereinafter
  referred to as ``Majority'';
\item the ``Low rank approximation'' decision rule applied to the ``Uniform allocation'' strategy,
  in the figures referred to as ``LRA uniform'';
\item the ``Low rank approximation'' decision rule applied to the ``Greedy allocation'' strategy, in the
  figures referred to as ``LRA greedy'';
\item the ``MAP'' decision rule applied to the ``Greedy allocation'' strategy, in the following referred
  to as ``MAP greedy''.
\end{itemize}
Specifically, for the greedy allocation algorithm, described in Section~\ref{sec:greedy}, we employed the
overall mutual information  $P_3(\Dc)$  as objective function.

The first set of results is reported in Figure~\ref{fig:figure1}. There we considered the most
classical scenario where tasks are identical.

The results depicted in Figure~\ref{fig:figure1}(a) assume that all workers belonging to the same class
have the same error probability i.e., $p_{tw}=\pi_{tk}$. In particular, we set $\pi_{t1}=0.1,
\pi_{t2}=0.2, \pi_{t3}=0.5$ for all $t$. This means that workers in class 1 are the most reliable,
while workers in class 3 are spammers.  Moreover, the number of available workers per class is set
to $W_1 = 30, W_2 = 120, W_3 = 150$.  The figure shows the average error probability achieved on the
tasks, plotted versus the average number of workers per task, $\beta=C/T$.  As expected, greedy
allocation strategies perform better due to the fact that they exploit the knowledge about the
workers' reliability ($p_{tw}$), and thus they assign to tasks the best possible performing workers.
These strategies provide quite a significant reduction of the error probability, for a given number of
workers per task, or a reduction in the number of assignments required to achieve a fixed target
error probability.  For example, $P_e=10^{-2}$ can be achieved by greedy algorithms by assigning only
4 workers per task (on average), while algorithms unaware of workers reliability require more than 20
workers per task (on average).  We also observe that the LRA algorithm proposed in~\cite{Devavrat}
performs similarly to the optimal MAP algorithm.

Next, we take into account the case where in each class workers do not behave exactly the same.  As
already observed, this may reflect both possible inaccuracies/errors in the reconstruction of user
profiles, and the fact that the behavior of workers is not fully predictable, since it may vary over time.  Specifically, we assume that, in each class, two types of workers coexist, each characterized
by a different error probability $p_{tw}$. More precisely, workers of type 1 have error
probability $p_{tw} = (1-x)\pi_{tk}$, while workers of type 2 have error probability probability
$p_{tw} = (1-x) \pi_{tk} +x/2$, where $0\le x\le 1$ is a parameter. Moreover workers are of type 1
and type 2 with probability $1-2\pi_{tk}$ and $2\pi_{tk}$, respectively, so that the average
error probability over the workers in class $k$ is $\pi_{tk}$.  We wish to emphasize that this
bimodal worker model, even if it may appear somehow artificial, is attractive for the following two
reasons: i) it is simple (it depends on only one scalar parameter $x$), and ii) it encompasses as
particular cases the two extreme cases of full knowledge and hammer-spammer.  Indeed, for $x=0$ all
workers in each class behave exactly the same (they all have error probability
$p_{tw}=p_{tk}$); this is the case depicted in Figure~\ref{fig:figure1}(a), while for $x=1$ we
recover the hammer-spammer scenario.  This case is represented in Figure~\ref{fig:figure1}(b), where
workers are spammers with probability $2\pi_{tk}$ and hammers with probability $1-2\pi_{tk}$.  Here,
the greedy allocation algorithms still outperform the others. However, the MAP decision rule
provides performance lower than the ``LRA greedy'' due to the following two facts: i) MAP adopts a
mismatched value of the error probability of individual workers, when $x\neq 0$,  ii) MAP does not
exploit the extra information on individual worker reliability that is possible to gather by jointly
decoding different tasks.
In Figure~\ref{fig:figure1}(c), for $\beta=10$, we show the error probability plotted versus the
parameter $x$. We observe that the performance of the ``MAP greedy'' strategy is independent on the
parameter $x$ while the performance of ``LRA greedy'' improve as $x$ increases. This effect can
be explained by observing that the LRA ability of distinguishing good performing workers from bad
performing workers within the same class increases as $x$ increases.

Next, we assume that the $T=100$ tasks are divided into 2 groups of 50 each. Workers processing
tasks of group 1 and 2 are characterized by average error probabilities $\pi_{t1}=0.05,
\pi_{t2}=0.1, \pi_{t3}=0.5$ and $\pi_{t1}=0.1, \pi_{t2}=0.2, \pi_{t3}=0.5$, respectively.  This
scenario reflects the case where tasks of group 2 are more difficult to solve than tasks of group 1
(error probabilities are higher). Workers of class $\Cc_3$ are spammers for both kinds of tasks.
The error probabilities provided by the algorithms under study are depicted in
Figure~\ref{fig:figure2}, as a function of $x$ and for $\beta=10$.

\insertfig{0.8}{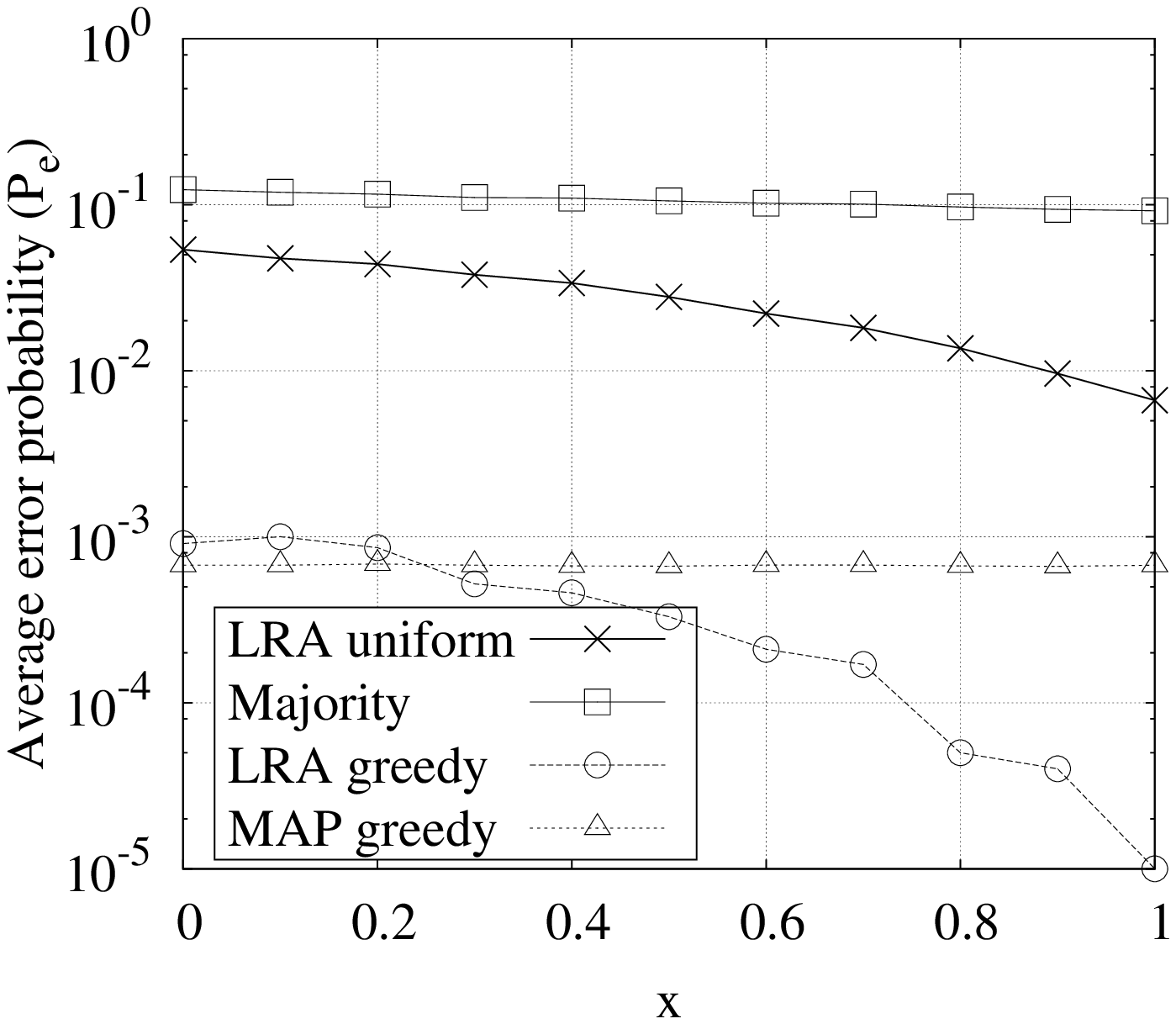}{Average error
  probability plotted versus $x$, for $\beta=10$ and task organized in two groups of different
  difficulties. For the first group of tasks $\pi_{t1}=0.05, \pi_{t2}=0.1,
\pi_{t3}=0.5$, while for the second group $\pi_{t1}=0.1, \pi_{t2}=0.2, \pi_{t3}=0.5$, Moreover $W_1 = 30,
W_2 = 120, W_3 = 150$.}{fig:figure2}

We observe that all strategies perform similarly, like in the scenario represented by
Figure~\ref{fig:figure1}(c), meaning that the algorithms are robust enough to deal with changes in
the behavior of workers with respect to tasks.  We wish to remark that the LRA decoding
scheme is fairly well performing also in this scenario, even if it was conceived and proposed only
for the simpler scenario of indistinguishable tasks.  This should not be too surprising, in light of
the fact that, even if the error probability of each user depends on the specific task, the relative
ranking among workers remains the same for all tasks.

\begin{figure*}[ht]
 \centering
 \subfigure[]
   {\includegraphics[width=0.32\textwidth]{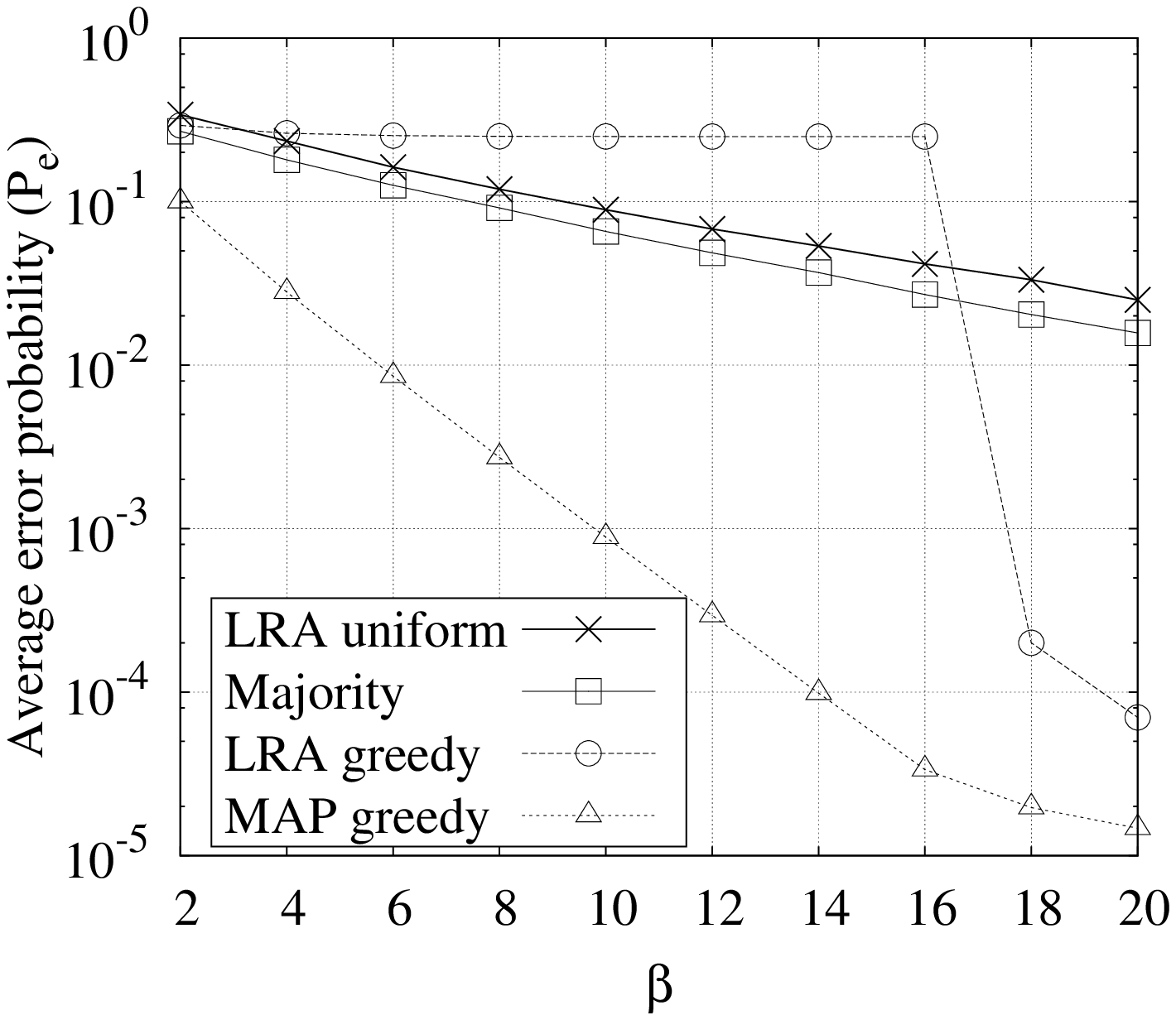}}
 \subfigure[]
   {\includegraphics[width=0.32\textwidth]{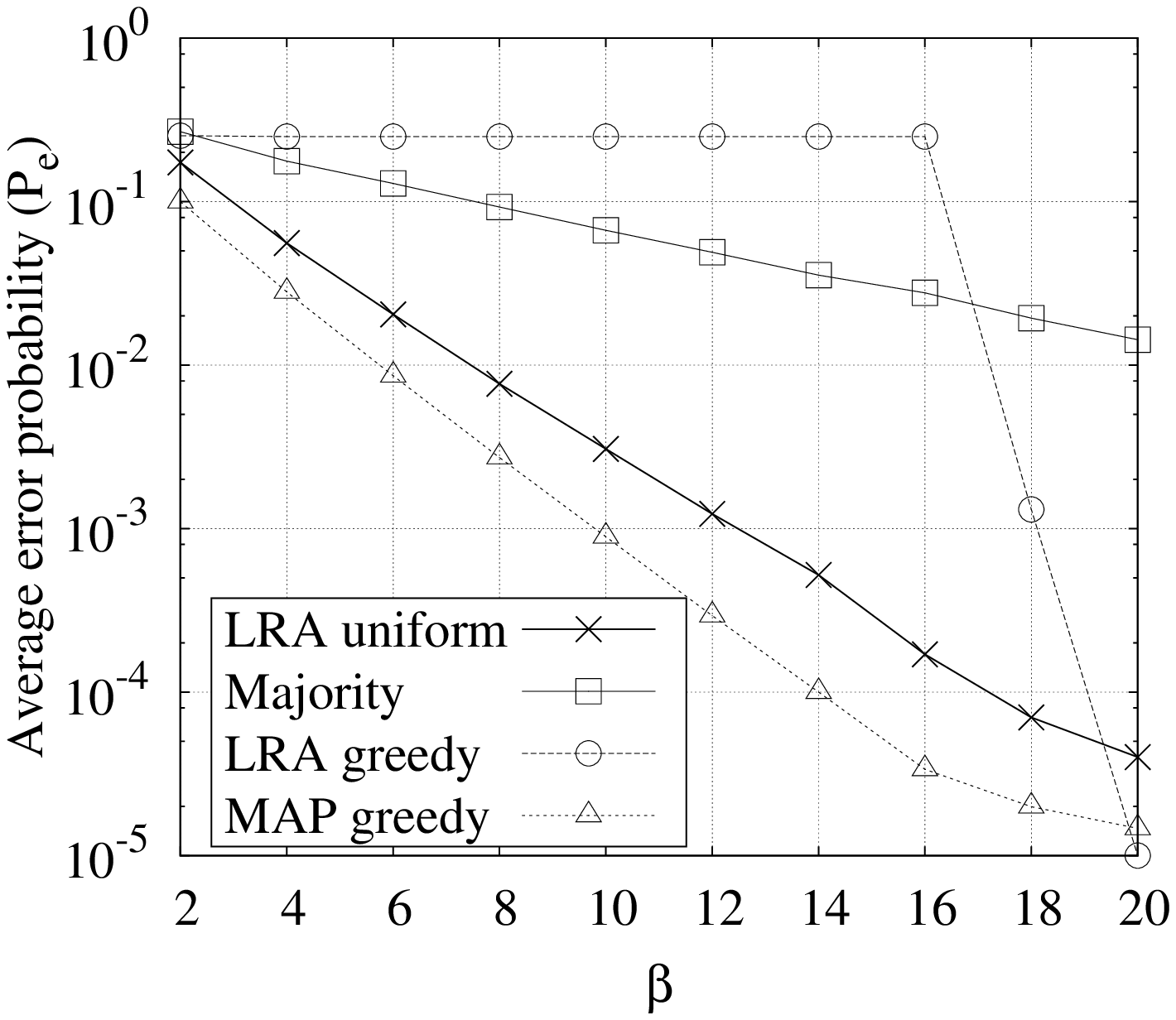}}
 \subfigure[]
   {\includegraphics[width=0.32\textwidth]{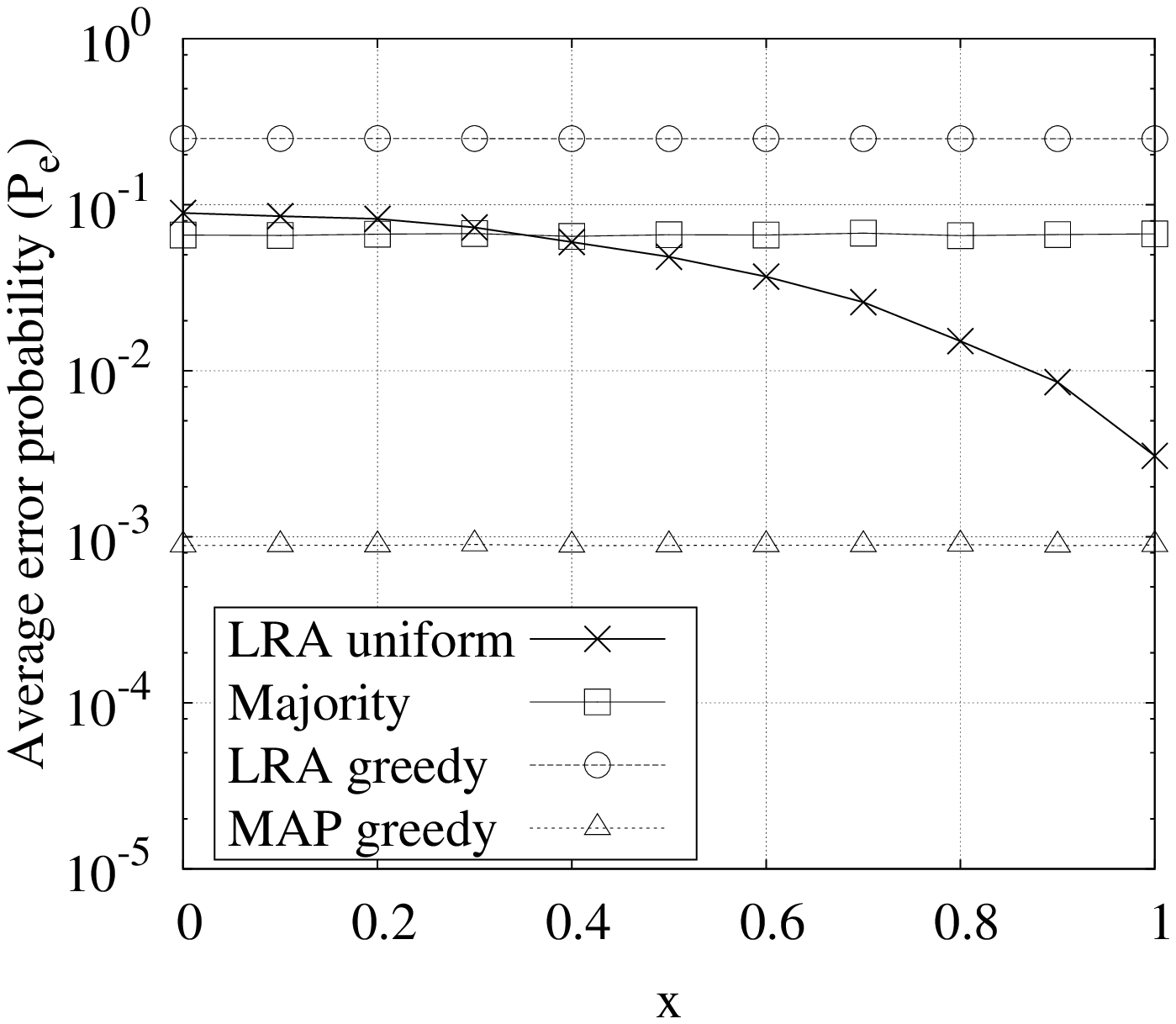}}
 \caption{Average error probability as a function of the average number of workers per task,
   $\beta$, and of the parameter $x$. Task organized in two groups of different difficulties. For
   the first group of tasks $\pi_{t1}=0.1, \pi_{t2}=0.25, \pi_{t3}=0.5$, while for the second group
   $\pi_{t1}=0.5, \pi_{t2}=0.25, \pi_{t3}=0.1$. Moreover $W_1 = 40, W_2 = 120$ and $W_3 = 40$. In figures
   (a) and (b) we set $x=0$ and $x=1$, respectively, while in figure (c) we set $\beta=10$.}
 \label{fig:figure3} 
\end{figure*}

Finally, in Figure~\ref{fig:figure3} we consider the same scenario as in
Figure~\ref{fig:figure2}. Here, however, the number of available workers per class is set to $W_1 =
40, W_2 = 120, W_3 = 40$, and the workers error probabilities for the tasks in group 1 and 2 are
given by $\pi_{t1}=0.1, \pi_{t2}=0.25,\pi_{t3}=0.5$, and $\pi_{t1}=0.5, \pi_{t2}=0.25,
\pi_{t3}=0.1$, respectively. This situation reflects the case where workers are more specialized or
interested in solving some kinds of tasks. More specifically, here workers of class 1 (class 3) are
reliable when processing tasks of group 1 (group 2), and behave as spammers when processing tasks of
group 2 (group 1). Workers of class 2 behave the same for all tasks.  In terms of performance, the
main difference with respect to previous cases is that the ``LRA greedy'' algorithm shows severely
degraded error probabilities for $\beta \le 16$.  This behavior should not surprise the reader,
since our third scenario may be considered as adversarial for the LRA scheme, in light of the fact
that the relative ranking among workers heavily depends on the specific task.  Nevertheless, it may
still appear amazing that ``LRA greedy'' behaves even worse than the simple majority scheme in several
cases.  The technical reason for this behavior is related to the fact that, in our example, for
$\beta \le 16$, tasks of group 1 (group 2) are allocated to workers of class 1 (class 3) only, whilst
workers of class 2 are not assigned any task. For this reason, the matrix $\Am$ turns out to have a
block diagonal structure, which conflicts with the basic assumption made by LRA that matrix $\EE
[\Am]$ can be well approximated by a unitary rank matrix. For $\beta > 16$, tasks are also allocated
to workers of class 2; in this situation, the matrix $\Am$ is not diagonal anymore, and the ``LRA
greedy'' performs quite well. Observe, however, that also in this case even a fairly imprecise
characterization of the worker behavior can be effectively exploited by the requester to
significantly improve system performance.



Finally, we want to remark that we have tested several versions of greedy algorithms under different
objective functions, such as $P_1(\Dc)$, $P_2(\Dc)$, and $P_3(\Dc)$, finding that they provide, in
general, comparable performance. The version employing mutual information was often providing
slightly better results, especially in the case of LRA greedy. This can be attributed to the
following two facts: i) the mutual information was proved to be submodular; ii) being mutual
information independent from the adopted decoding scheme, it provides a more reliable metric for
comparing the performance of different task allocations under the LRA decoding scheme with respect
to the error probability $P_1(\Dc)$ (which, we recall, is computed under the assumption that the
decoding scheme is MAP).  Unfortunately, due to the lack of space, we cannot include these results
in the paper.

\section{Concluding Remarks}
\label{sec:conclusions}

In this paper we have presented the first systematic investigation of the impact of information
about workers' reputation in the assignment of tasks to workers in crowdsourcing systems,
quantifying the potential performance gains in several cases.  We have formalized the optimal task
assignment problem when workers' reputation estimates are available, as the maximization of a
monotone (submodular) function subject to Matroid constraints. Then, being the optimal problem
NP-hard, we have proposed a simple but efficient greedy heuristic task allocation algorithm,
combined with a simple ``maximum a-posteriori`` decision rule.  We have tested our proposed
algorithms, and compared them to different solutions, which can be obtained by extrapolating the
proposals for the cases when reputation information is not available, showing that the crowdsourcing
system performance can greatly benefit from even largely inaccurate estimates of workers'
reputation.
Our numerical results have shown that:
\begin{itemize}
 \item even a significantly imperfect characterization of the workers' earnestness can be extremely
   useful to improve the system performance;
\item the application of advanced joint tasks decoding schemes such as LRA can further improve the
  overall system performance, especially in the realistic case in which the a-priori information
  about worker reputation is largely affected by errors;
\item the performance of advanced joint tasks decoding schemes such as LRA may become extremely poor
  in adversarial scenarios.
\end{itemize}

\appendices
 
\section{Mutual information for known error probabilities $\pi_{tk}$ \label{app:mutual}}
The workers' answers about the tasks $\tauv$ are collected in the random $T\times W$ matrix $\Am$,
defined in Section~\ref{sec:SA}.  The information that the answers $\Am$ provide about the tasks
$\tauv$ is denoted by
\[ \Ic(\Am;\tauv) = H(\Am)- H(\Am|\tauv) \]
where the entropy $H(a)$ and the conditional entropy $H(a|b)$ have been defined in
Section~\ref{sec:mutual_info}.  We first compute $H(\Am|\tv)$ and we observe that, given the tasks
$\tauv$, the answer $\Am$ are independent, i.e., $\PP(\Am | \tauv)= \prod_{k=1}^K\prod_{t=1}^T \PP(\av_{tk}|\tau_t)$,
where $\av_{tk}$ is the vector of answers to task $\theta_t$ from users of class $\Cc_k$. 
Since $\PP(\Am | \tauv)$ has a product form, we obtain $H(\Am|\tauv) = \sum_{k=1}^K\sum_{t=1}^T H(\av_{tk}|\tau_t)$.
Thanks to the fact that workers of the same class are independent and all have error 
probability $\pi_{tk}$, we can write
$H(\av_{tk}| \tau_t) = d_{ik}H_b(\pi_k)$ where $H_b(p) = -p\log p -(1-p)\log(1-p)$ and
$d_{tk}$ is the number of allocations of task $t$ in class $\Cc_k$.  In conclusion, we get:
\[  H(\Am|\tauv) = \sum_{t=1}^T \sum_{k=1}^K  d_{tk} H_b(\pi_{tk})\]

As for the entropy $H(\Am)$, we have:
\[ \PP(\Am)  =\EE_{\tauv}\PP(\Am|\tauv) = \EE_{\tauv}\prod_{t=1}^T \PP(\av_t | \tau_t) = \prod_{t=1}^T \EE_{\tau_t}\PP(\av_{t} | \tau_t)\]
 where $\av_t$ is the vector of answers to task $\theta_t$ (corresponding to the $t$-th row of
 $\Am$).  Note that $\EE_{\tau_t}\PP(\av_{t} | \tau_t)=\PP(\av_t)$, hence $\PP(\Am)=\prod_{t=1}^T
 \PP(\av_t)$ and we immediately obtain $H(\Am) = \sum_{t=1}^T H(\av_t)$.  The probabilities
 $\PP(\av_{tk} | \tau_t=1)$ and $\PP(\av_{tk} | \tau_i=-1)$ are easy to compute.  Indeed for
 $\tau_t=-1$ we have
\begin{equation}
  \PP(\av_{tk} | \tau_t=-1) = \pi_{tk}^{d_{tk}-m_{tk}}(1-\pi_{tk})^{m_{tk}}
  \label{eq:p_y|t0}
\end{equation}
where $m_{tk}$ is the number of ``$-1$'' answers to task $\theta_t$ from class-$k$ workers.  The above
formula derives from the fact that workers of the same class are independent and have the same
error probability $\pi_{tk}$.  Similarly
\begin{equation}
  \PP(\av_{tk} | \tau_t=+1) = \pi_{tk}^{m_{tk}}(1-\pi_{tk})^{d_{tk}-m_{tk}}
  \label{eq:p_y|t1}
\end{equation}
The expressions~\eqref{eq:p_y|t0} and~\eqref{eq:p_y|t1} can compactly written as
\begin{equation}
  \PP(\av_{tk} | \tau_t) = (1-\pi_{tk})^{d_{tk}} b_{tk}^{(1-\tau_t)d_{tk}/2-m_{tk}\tau_t} 
  \label{eq:p_y|t}
\end{equation}
where $b_{tk}=\pi_{tk}/(1-\pi_{tk})$. Since, given $\tau_t$, workers are independent, we obtain
\begin{eqnarray}
\PP(\av_t)
&=&\EE_{\tau_t}\PP(\av_{t} | \tau_t) \non
&=&  \gamma_{tk}\EE_{\tau_t}\left[\prod_{k=1}^K b_{tk}^{(1-\tau_t)d_{tk}/2-m_{tk}\tau_t} \right] \non
&=&  \frac{\gamma_{tk}}{2} \left[\prod_{k=1}^K b_{tk}^{-m_{tk}} + \prod_{k=1}^K b_{tk}^{d_{tk}+m_{tk}} \right] =  \frac{\gamma_{tk}}{2} f(\mv_{t}) \nonumber
\end{eqnarray}
with $\mv_{t} = [m_{t1},\ldots,m_{tK}]$ and $f(\nv) =\prod_{k=1}^K b_{tk}^{-n_{k}} + \prod_{k=1}^K b_{tk}^{d_{tk}+n_{k}}$.
Finally, by using the definition of entropy,
\begin{eqnarray} 
H(\av_t) 
&=& \EE_{\av_t}[-\log \PP(\av_t)] \non
&=& -\log\frac{\gamma_{tk}}{2} -\EE_{\av_t}f(\mv_{t}) \non
&=& -\log\frac{\gamma_{tk}}{2} -\frac{\gamma_{tk}}{2}  \sum_{\nv} f(\nv)\log f(\nv)\prod_{k=1}^K\binom{m_{tk}}{n_k}
\nonumber
\end{eqnarray}
where $\nv=[n_{1},\ldots,n_{K}]$ and $n_k=0,\ldots,m_{tk}$, $k=1,\ldots,K$. 

\section{Matroid definition and Proof of Proposition  \ref{prop-Matroid}  \label{app:matroid}}

First we recall the definition of a Matroid. Given a family $\Fb$ of subsets of a finite ground set
$\Oc$ (i.e., $\Fb \subset 2^{\Oc})$ $\Fb$ is a Matroid iff: i) if $\Gc\in \Fb $ then $\Hc \in \Fb$
whenever $\Hc \subseteq \Gc$\\ ii) if $\Gc\in \Fb $ and $\Hc \in \Fb$ with $ |\Gc|>|\Hc|$ then an
$(t_0,w_0) \in \Gc \setminus \Hc$

Now we can prove Proposition~\ref{prop-Matroid}.  First observe that in our case property i)
trivially holds. Now we show that property ii) holds too.  Given that $ |\Gc|>|\Hc|$ and since by
construction $ |\Gc| =\sum_w \Lc(w,\Gc)$ and $ |\Hc| =\sum_w \Lc(w,\Hc)$, necessarily there exists
an $w_0$ such that $|\Lc(w_0,\Gc)|>|\Lc(w_0,\Hc)|$, this implies that $\Lc(w_0,\Gc) \setminus
\Lc(w_0,\Hc)\neq \emptyset$.  Let $(t_0, w_0)$ be an individual assignment in $\Lc(w_0,\Gc)
\setminus \Lc(w_0,\Hc)$.  Since by assumption $|\Lc(w_0,\Hc)|<|\Lc(w_0,\Gc)|\le r_{w_0}$, denoted
with $\Hc'= \Hc \cup \{(t_0,w_0)\}$, we have that $|\Lc(w_0,\Hc')|=|\Lc(w_0,\Hc)| +1 \le
|\Lc(w_0,\Gc)|\le r_{w_0}$ similarly $|\Hc'|=|\Hc|+1\le |\Gc|\le C$, therefore $\Hc' \in \Fb$.

The fact that in our case $q=\frac{\max_{\Gc in \Bb}  |\Gc|}{\min_{\Gc in \Bb}  |\Gc|}=1$ descends immediately by the fact that necessarily $\Gc \in \Bb$ 
iff either i)  $|\Gc|=C$   when    $C\le \sum_w r_w$   or  ii) $ |\Gc|= \sum_w r_w$  when   $C> \sum_w r_w$. 
 
\section{Submodularity of the mutual information\label{app:submodularity}}

Let $\Gc_1$ and $\Gc_2$ be two generic allocations for the task $\tau$, such that $\Gc_2 \subseteq
\Gc_1$ and $\Gc_1 = \Gc_2 \cup \Gc_3$. Also let the pair $(t,w)\in \Oc \setminus \Gc_1$.  With a
little abuse of notation hereinafter we denote by $\Ic(\Gc,\tau)$ the mutual information between the
task $\tau$ and the vector of answers $\av(\Gc)$. Then the mutual information $I(\Gc; \tau)$ is submodular if
\begin{equation}
I(\Gc_2 \cup (t,w); \tau)-I(\Gc_2; \tau) \ge I(\Gc_1 \cup (t,w); \tau)-I(\Gc_1; \tau) 
\label{eq:submodularity_I}
\end{equation}
We first observe that
\begin{eqnarray*}
I(\Gc_1 \cup (t,w); \tau) 
&=& I(\Gc_2 \cup \Gc_3 \cup (t,w); \tau) \non
&\stackrel{(a)}{=}& I(\Gc_2\cup (t,w); \tau) + I(\Gc_3; \tau | \Gc_2\cup (t,w))
\end{eqnarray*}
where in $(a)$ we applied the mutual information chain rule.
Similarly we can write $I(\Gc_1; \tau) = I(\Gc_2; \tau) + I(\Gc_3; \tau | \Gc_2)$
By consequence ~\eqref{eq:submodularity_I} reduces to
\[ I(\Gc_3; \tau | \Gc_2) \ge I(\Gc_3; \tau | \Gc_2\cup (t,w)) \]
By using the definition of the mutual information given in \eqref{eq:mutual_info_definition}
we obtain 
\begin{equation} H(\Gc_3 | \Gc_2) -  H(\Gc_3 | \tau, \Gc_2) \ge H(\Gc_3 | \Gc_2\cup (t,w)) -H(\Gc_3 |\tau, \Gc_2\cup (t,w))
\label{eq:submodularity_I2}
\end{equation}

We now observe that $H(\Gc_3 | \tau, \Gc_2) = H(\Gc_3 |\tau, \Gc_2\cup (t,w)) = H(\Gc_3 | \tau)$
since $\Gc_3$ only depends on $\tau$. Therefore ~\eqref{eq:submodularity_I2} reduces to
\[ H(\Gc_3 | \Gc_2) \ge H(\Gc_3 | \Gc_2\cup (t,w))  \]
which holds due to the fact that entropy is reduced by conditioning.

\end{document}